\documentstyle[pasjskel,graphicx,natbib,twocolumn]{article}

\begin{document}
\title{Changing Supercycle of the ER UMa-Type Star V1159 Ori}

\author{Taichi \textsc{Kato}}
\affil{Department of Astronomy, Faculty of Science, Kyoto University,
       Sakyo-ku, Kyoto 606-8502}
\email{tkato@kusastro.kyoto-u.ac.jp}

\begin{abstract}
   We examined the VSNET light curve of the ER UMa-type star V1159 Ori.
We detected a large variation of the supercycle (the interval between
successive superoutbursts) between extremes of 44.6 and 53.3 d.
The outburst activity was also found to decrease when the supercycle
was long.  The observed variation of the supercycle corresponds to a
variation of $\sim$40\% of the mass-transfer rate from the secondary star,
totally unexpected for this class of objects.  We also detected a hint
of $\sim$1800 d periodicity in the variation, whose period is close to
what has been suggested for solar-type cycles for cataclysmic variables
(CVs).  If this periodicity is caused by the magnetic activity of the
secondary star, this detection constitutes the first clear evidence of
continuing magnetic activity in CV evolution, even after crossing the
period gap.  This activity may partly explain still poorly understood
origins of the high mass-transfer rates in ER UMa-type stars.
\end{abstract}


\section{Introduction}
   ER UMa stars are a subgroup of SU UMa-type dwarf novae (for a review
of dwarf novae, see \cite{osa96}), whose known members are ER UMa, V1159 Ori,
RZ LMi, and DI UMa.  The most striking feature of ER UMa stars is the
extremely short recurrence time (19--45 d) of superoutbursts (\cite{kat95};
\authorcite{nog95a} 1995a,b; \cite{rob95}; \cite{mis95};
\cite{kat96}).
Another striking feature of ER UMa stars is the stability of supercycles,
both in their lengths and outburst pattern.  The best exemplification of
this stability can be seen in folded light curves and $O-C$ figures presented
in \citet{rob95}.  The extremely short supercycle length and
the stability of the outburst patterns are basically explained, within the
framework of the disk-instability model, as a result of constant high
mass-transfer rates from the secondary \authorcite{osa95a}
(\yearcite{osa95a}).  The mass-transfer
rates in SU UMa-type dwarf novae are generally considered to be confined
to a small range determined by angular-momentum removal by the
gravitational wave radiation.  The origin of high-mass transfer rates in
ER UMa stars is still an open question.  Some models assume irradiation
effect from a hot white dwarf, which may be the result of a hypothetical
recent nova eruption (the possibility was originally raised by
\authorcite{nog95b} \yearcite{nog95b}, see also \cite{pat98}).
An examination of any secular changes in the supercycle in
these systems would provide an essential clue for testing these hypotheses.

\section{Observation and Analysis}
\begin{figure*}
  \begin{center}
    \FigureFile(180mm,120mm){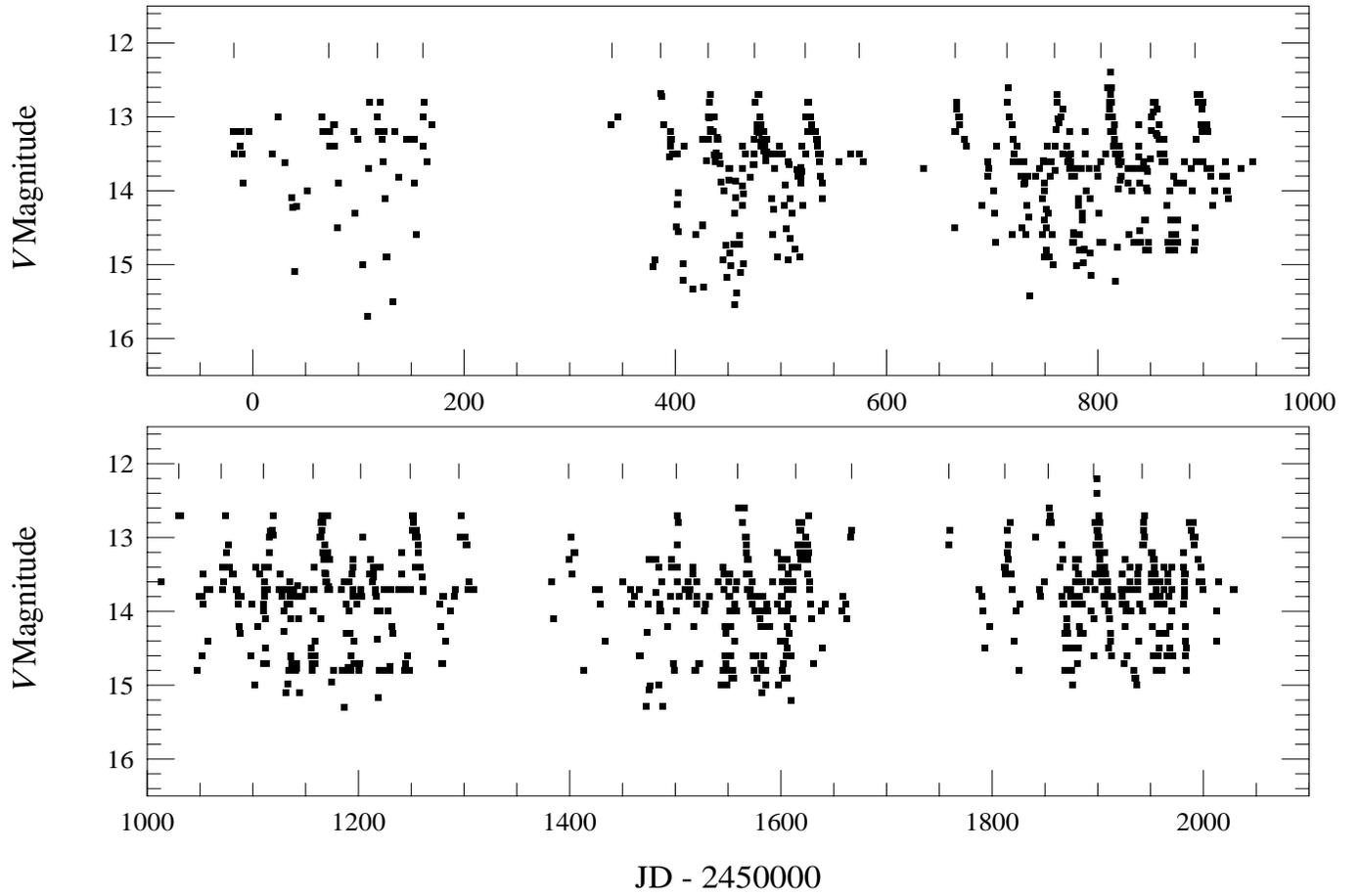}
  \end{center}
  \caption{Light curve of V1159 Ori from VSNET observations.  The ticks
  represent the start of superoutbursts, as listed in table
  \ref{tab:table1}}\label{fig:figure1}.
\end{figure*}

   We examined the observations posted to VSNET
{\tt $\langle$http://www.kusastro.kyoto-u.ac.jp/vsnet/$\rangle$},
and found an appreciable change in one of the ER UMa stars, V1159 Ori.
The object has been very well sampled by many observers around the world
since 1995 September (figure \ref{fig:figure1}).

The time of the start of a superoutburst was defined as its mid-rising
branch.  Although occasional observational gaps introduced an uncertainty
of 1--2 d, most of these superoutbursts were well sampled and
the times were usually determined within an uncertainty of 1 d.  Table
\ref{tab:table1} lists the observed times of superoutbursts.  The cycle
number ($E$) represents number of supercycles since the JD 2449982
superoutburst.

\begin{table}
  \caption{Superoutbursts of V1159 Ori.}\label{tab:table1}
  \begin{center}
    \begin{tabular}{cccc}
    \hline\hline
    JD start & Cycle number & JD start & Cycle number \\
    \hline
    2449982 &  0 & 2451110 & 25 \\
    2450072 &  2 & 2451157 & 26 \\
    2450118 &  3 & 2451202 & 27 \\
    2450161 &  4 & 2451249 & 28 \\
    2450340 &  8 & 2451295 & 29 \\
    2450386 &  9 & 2451399 & 31 \\
    2450431 & 10 & 2451450 & 32 \\
    2450475 & 11 & 2451501 & 33 \\
    2450523 & 12 & 2451559 & 34 \\
    2450574 & 13 & 2451614 & 35 \\
    2450665 & 15 & 2451667 & 36 \\
    2450714 & 16 & 2451759 & 38 \\
    2450759 & 17 & 2451812 & 39 \\
    2450803 & 18 & 2451853 & 40 \\
    2450850 & 19 & 2451896 & 41 \\
    2450892 & 20 & 2451942 & 42 \\
    2451030 & 23 & 2451987 & 43 \\
    2451070 & 24 &         &    \\
    \hline
    \end{tabular}
  \end{center}
\end{table}

\vskip 3mm

\begin{figure}
  \begin{center}
    \FigureFile(80mm,60mm){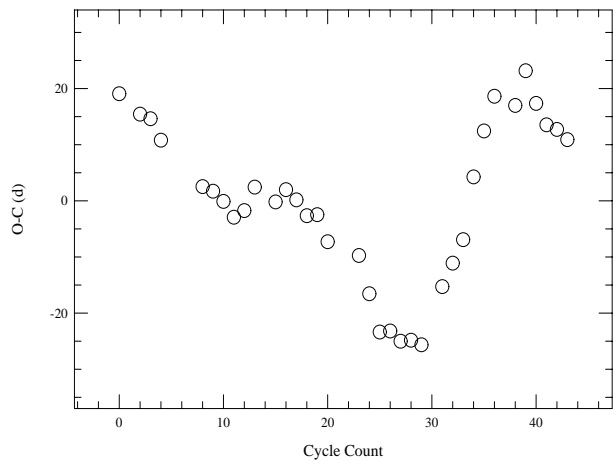}
  \end{center}
  \caption{$O-C$ diagram of V1159 Ori superoutbursts.}\label{fig:figure2}
\end{figure}

   A regression to these times has yielded a linear ephemeris of
$2449962.9 + 46.82 E$.  The derived supercycle length of 46.82 d is
slightly longer than the 44.5 d by \citet{rob95}.
Figure \ref{fig:figure2} shows the $O-C$ diagram against this ephemeris.
The most remarkable feature is the presence of large $O-C$ changes compared
to \citet{rob95}.  This large change is mainly caused by
an increase in the supercycle length between $E=29$ and $E=36$, corresponding
to the period between 1999 May and 2000 May.  The supercycle during
interval is 53.3 d, which is 14\% longer than the long-term average.
Such a large change in supercycle has not been seen in ER UMa.

\begin{figure*}
  \begin{center}
    \FigureFile(140mm,90mm){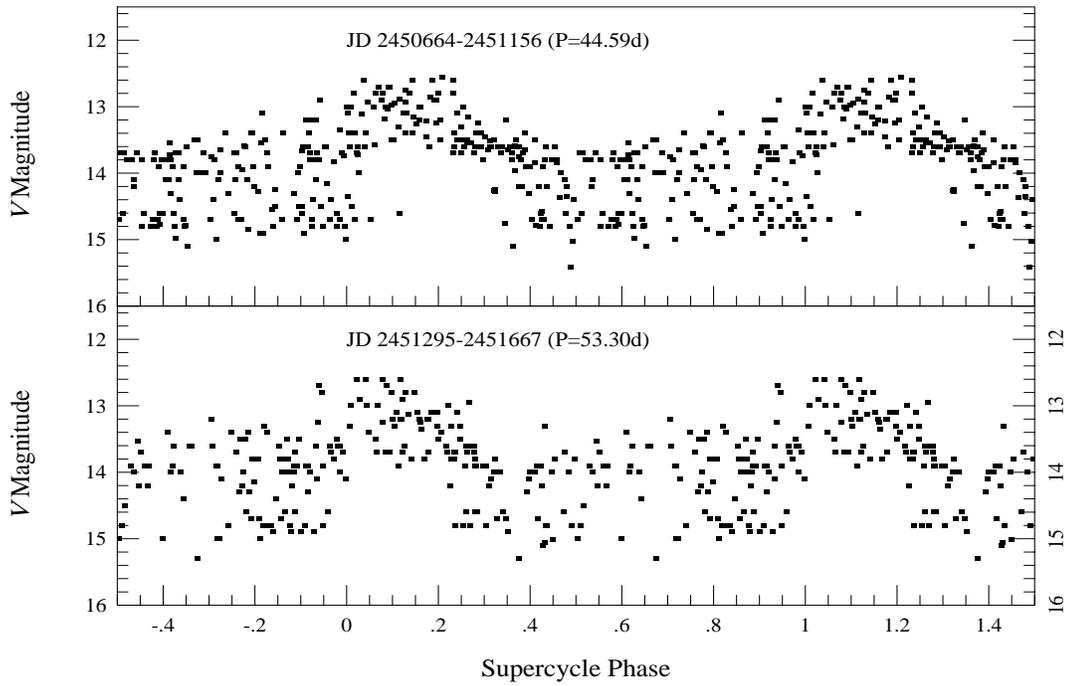}
  \end{center}
  \caption{Folded light curves.  The upper panel shows the epoch with
  a short supercycle length (44.59 d).  The lower panel shows the epoch
  with a long supercycle (53.30 d) and decreased outburst activity.
  The duty cycle of a superoutburst (phase 0--0.45 in the upper panel,
  phase 0--0.35 in the lower panel) is markedly decreased in the latter
  epoch.}\label{fig:figure3}
\end{figure*}

\section{Discussion}
   The long-term average of the supercycle lengths in V1159 Ori, being
close to the minimum value predicted by \authorcite{osa95a}
(\yearcite{osa95a}), the supercycle length near this period is expected
to be insensitive to the mass-transfer rate from the secondary.
If the observed change in V1159 Ori was caused by the variable
mass-transfer rate, a relatively large change is necessary
to reproduce the observation.  Using the $\dot{M}-{\rm supercycle}$
diagram in \authorcite{osa95a} (\yearcite{osa95a}), a supercycle of
53.3 d corresponds to a reduction of $\sim$40\% of mass-transfer rates
from what is expected for a 44.5-d supercycle.  The marked reduction of
the superoutburst duty cycle during this period (figure \ref{fig:figure3})
also supports this interpretation.

   Another observational evidence of a large period change in ER UMa stars
has been reported in DI UMa \citep{fri99}.  However, the extreme shortness of
supercycles in DI UMa and RZ LMi requires an additional (still poorly
identified) mechanism (\authorcite{osa95a} \yearcite{osa95a}),
and its change may be of different nature.  Another noteworthy feature
in the observed $O-C$ diagram of V1159 Ori is a possible periodicity with
a period of $\sim$38 cycles,
corresponding to $\sim$1800 d, rather than a monotonous change originally
proposed by \citet{rob95}; this is contrary to the expected effect
by decreasing heating from a hypothetical recent nova eruption on
a white dwarf.  The observed possible long-term period is close to those
observed as possible solar-type cycles in cataclysmic variables
(e.g. \cite{bia88}; \cite{ak01}).  If such a ``solar-type" cycle is
responsible for the change in the supercycle of V1159 Ori, this may provide
promising evidence for the presence of magnetic activity in dwarf novae
below the period gap, which has usually been considered to cease, or to be
markedly reduced, when the secondary becomes fully convective after crossing
the period gap.  Furthermore, the continuing magnetic activity may be one
of the mechanisms for effectively removing the angular momentum from the
binary system, by which the required high mass-transfer in ER UMa-type
systems may be partly explained.

\vskip 3mm

The author is grateful to VSNET members, especially to Rod Stubbings,
Gene Hanson, Gary Poyner, Andrew Pearce, Seiichiro Kiyota, Eddy Muyllaert,
Tsutomu Watanabe and numerous observes for providing vital observations.


\begin{thebibliography}{}
\bibitem[Ak et al. (2001)]{ak01}
  Ak, T., Ozkan, M. T., \& Mattei, J. A.\ 2001, \aap, 369, 882
\bibitem[Bianchini (1988)]{bia88}
  Bianchini, A.\ 1988, Inf. Bull. Var. Stars, 3136
\bibitem[Fried et al. (1999)]{fri99}
  Fried, R. E., Kemp, J., Patterson, J., Skillman, D. R., Retter, A.,
  Leibowitz, E., \& Pavlenko, E.\ 1999, \pasp, 111, 1275
\bibitem[Kato, Kunjaya (1995)]{kat95}
  Kato, T., \& Kunjaya, C.\ 1995, \pasj, 47, 163
\bibitem[Kato et al. (1996)]{kat96}
  Kato, T., Nogami, D., \& Baba, H.\ 1996, \pasj, 48, L93
\bibitem[Misselt, Shafter (1995)]{mis95}
  Misselt, K. A., \& Shafter, A. W.\ 1995, \aj, 109, 1757
\bibitem[Nogami et al. (1995a)]{nog95a}
  Nogami, D., Kato, T., Masuda, S., \& Hirata, R.\ 1995a,
  Inf. Bull. Var. Stars, 4155
\bibitem[Nogami et al. (1995b)]{nog95b}
  Nogami, D., Kato, T., Masuda, S., Hirata, R.., Matsumoto, K., Tanabe, K., 
  \& Yokoo, T.\ 1995b, \pasj, 47, 897
\bibitem[Osaki (1995a)]{osa95a}
  Osaki, Y.\ 1995a, \pasj, 47, L11
\bibitem[Osaki (1995b)]{osa95b}
  Osaki, Y.\ 1995b, \pasj, 47, L25
\bibitem[Osaki (1996)]{osa96}
  Osaki, Y.\ 1996, \pasp, 108, 39
\bibitem[Patterson (1998)]{pat98}
  Patterson, J.\ 1998, \pasp, 110, 1132
\bibitem[Robertson et al. (1995)]{rob95}
  Robertson, J. W., Honeycutt, R. K., \& Turner, G. W.\ 1995, \pasp,
  107, 443
\end{thebibliography}
\end{document}